\documentclass[a4paper,11pt]{article}
\usepackage{pos}
\title{Study of the calibration method using the stars measured by the EUSO-TA telescope}
 \ShortTitle{EUSO-TA calibration}

\author*[a,b]{Z. Plebaniak}
\author[c]{M. Przybylak}
\author[a,b]{D. Barghini}
\author[a]{M. Bertaina}
\author[a,b]{F. Bisconti}
\author[d,e,f]{~~~~~~~~~~~~~~~~~~M. Casolino}
\author[b,g]{D. Gardiol}
\author[h]{R. Lipiec}
\author[i]{L. W. Piotrowski}
\author[c]{K. Shinozaki}
\author[c]{~~~~~~~~~~J. Szabelski}
\forColl{JEM-EUSO}

\affiliation[a]{Dipartimento di Fisica, Universit\`{a} di Torino, Italy}
\affiliation[b]{Instituto Nazionale di FisicaNucleare - Sezione di Torino, Italy}
\affiliation[c]{National Centre for Nuclear Research, {\L }\'od\'z, Poland}
\affiliation[d]{RIKEN, Wako, Japan}
\affiliation[e]{Instituto Nazionale di Fisica Nucleare - Sezione di Roma Tor Vergata, Italy}
\affiliation[f]{Universit\`{a} degli Studi di Roma Tor Vergata - Dipartimento di Fisica, Roma, Italy}
\affiliation[g]{Instituto Nazionale di Astrofisica - Osservatorio Astrofisico di Torino, Italy}
\affiliation[h]{{\L }\'od\'z University of Technology, {\L }\'od\'z, Poland}
\affiliation[i]{Faculty of Physics, University of Warsaw, Poland}

\emailAdd{zp@zpk.u.lodz.pl}
\emailAdd{mp@zpk.u.lodz.pl}

\abstract{EUSO-TA is a ground-based experiment, placed at Black Rock Mesa of the Telescope Array site as a part of the JEM-EUSO (Joint Experiment Missions for the Extreme Universe Space Observatory) program. 
The UV fluorescence imaging telescope with a field of view of about 10.6$^\circ$ x 10.6$^\circ$ consisting of 2304 pixels (36 Multi-Anode Photomultipliers, 64 pixels each) works with 2.5-microsecond time resolution. 
An experimental setup with two Fresnel lenses allows for measurements of Ultra High Energy Cosmic Rays in parallel with the TA experiment as well as the other sources like flashes of lightning, artificial signals from UV calibration lasers, meteors and stars. 
Stars increase counts on pixels while crossing the field of view as the point-like sources.
In this work, we discuss the method for calibration of EUSO fluorescence detectors based on signals from stars registered by the EUSO-TA experiment during several campaigns. 
As the star position is known, the analysis of signals gives an opportunity to determine the pointing accuracy of the detector.
This can be applied to space-borne or balloon-borne EUSO missions. 
We describe in details the method of the analysis which provides information about detector parameters like the shape of the point spread function and is the way to perform absolute calibration of EUSO cameras.}

\FullConference{37$^{\rm{th}}$ International Cosmic Ray Conference (ICRC 2021)\\
		July 12th -- 23rd, 2021\\
		Online -- Berlin, Germany}

\begin{document}
\maketitle

\section{Introduction}
EUSO-TA which is a part of the JEM-EUSO (Joint Experiment Missions - Extreme Universe Space Observatory) scientific program is a ground-based experiment located in Black Rock Mesa, the Telescope Array \cite{TAMEDA200974} (TA) experiment site. 
Experiment is designed to observe an Ultra High Energy Cosmic Rays (UHECR) simultaneously with TA. 
The detector is based on one Photo Detection Module (PDM) - the main unit of all EUSO experiments. 
PDM consists of an array of 3x3 Elementary Cells (ECs), each EC consists of 2x2 Hamamatsu R11265-M64 Multi-Anode Photomultipliers (MAPMT).
The total number of pixels on focal surface is 2304.
Each EC has separate  High Voltage Power Supply system \cite{Plebaniak:2017itg}.
Readout system is based on SPACIROC-1 ASIC \cite{AHMAD20121600}.
Applied optical system consist of two PMMA Fresnel lenses \cite{2013ICRC...33..627T} with effective area $\sim$0.92 $m^{2}$ allows for observations of the night sky in wide, 10.6$^\circ$ field of view.


\begin{figure}[!h]
\begin{center}
\includegraphics[width=.67\textwidth]{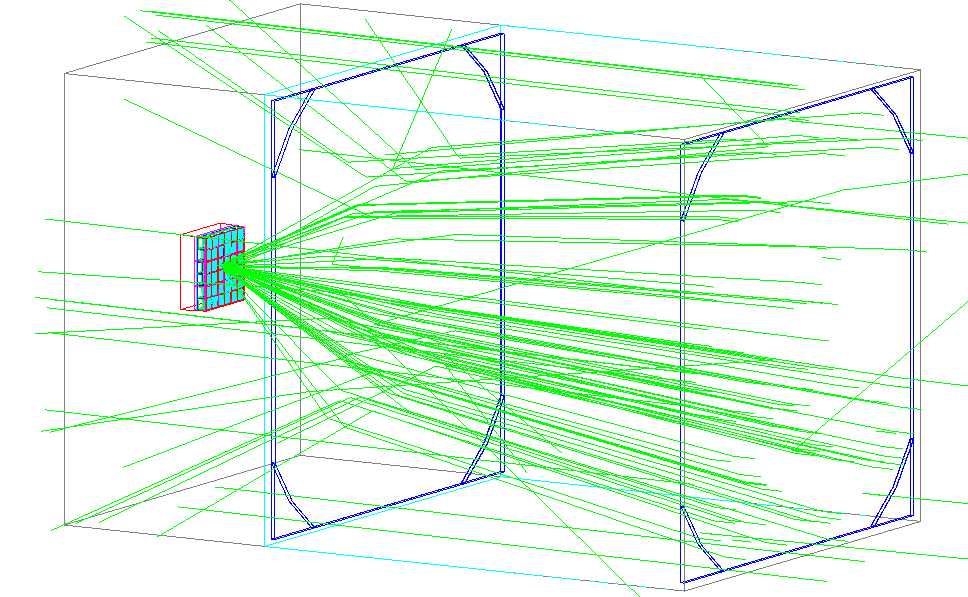}
\includegraphics[width=.72\textwidth]{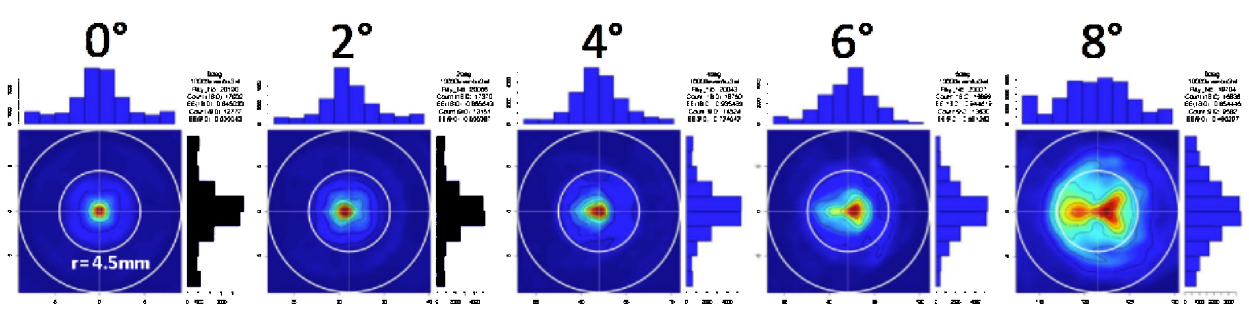}
	\caption{An example of visualization of the EUSO-TA ray-tracing simulation performed in the OffLine framework (top) \cite{Paul:2017scv} and PSF spot diagrams for incident angles between 0$^\circ$ and 8$^\circ$ (bottom) \cite{2013ICRC...33..627T}}
\label{OpticsEUSOTA}
\end{center}
\end{figure}

Since installation in 2013 several measurement sessions with TA have been conducted. 
During measurements artificial and non artificial sources of UV signal including UHECR, meteors, stars and calibration lasers have been observed.  
More detailed description of these observations and of the detector configuration can be found in \cite{Abdellaoui:2018rkw}.   
Analysis of the taken data and its comparison with TA gives us great possibility to test and improve EUSO technology - what is one of the main objectives of the EUSO-TA experiment.

Absolute calibration of EUSO detectors is an important step necessary to understand registered signal.
EUSO-TA fluorescence detector is sensitive to the UV light from stars. 
Photometry of stars is a standard method commonly used for calibration of optical telescopes. 
In this work we present procedure and results of absolute calibration of EUSO-TA detector based on observed UV signals from stars.

\section{The EUSO-TA detector}

The EUSO-TA fluorescence detector is located in front of TA-FD station at Black Rock Mesa site (1400m a.s.l.).
Elevation of the optical axis can be set manually in the rage between 0 and 30 degrees with respect to the horizon.
Focal surface containing the matrix of 48x48 pixels consists of the 64-channel MAPMTs with $\sim$75$\%$ collection efficiency and quantum efficiency with maximum reaching 35$\%$ at $\sim$350 nm.
Single pixels are squares with a size of 2.88 mm corresponding to 0.2$^{\circ}$ x 0.2$^{\circ}$ field of view. 
The nominal wavelength range for UHECR observations is 290 - 400 nm however signals from stars can be observed up to 600 nm due to transmittance of applied BG3 filter.
The photon collection efficiency of the EUSO-TA taking into account the surface reflection, PMMA absorption and other loss factors indicates wavelength-dependence with average value $\sim$40$\%$ for 0$^{\circ}$ of the incident angle \cite{2013ICRC...33..627T}.
Measured PSF for 0$^{\circ}$ gives around 80$\%$ of signal from point-like source in 3x3 pixels. This value is decreasing with the distance from optical axis.   


\begin{figure}[!h]
\begin{center}
\includegraphics[width=.73\textwidth]{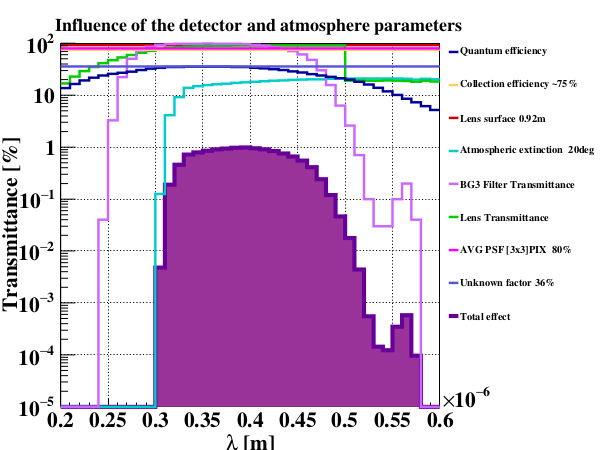}
	\caption{Known parameters of EUSO-TA detector. Blue, red and green lines represents wavelength-dependent parameters of EUSO-TA detector while violet line is related to total apparatus effect including PMT collection efficiency and ASIC inefficiency due to pileup depends on light intensity.}
\label{DetParam}
\end{center}
\end{figure}

SPACIROC1 front-end ASIC allows for work in single photon counting mode with 30ns of double pulse resolution.
Counts are sampled in 2.5 $\mu$s gate time units (GTUs) with 0.2 $\mu$s dead time.

Relation between registered $PE_{R}$ and generated $PE_{G}$ number of photo-electrons can be expressed in the following way \cite{Abdellaoui:2019qmg}:
\begin{equation}
	PE_{R} = PE_{G} \cdot \exp\left(-\frac{PE_{G} \cdot \tau}{T_{acq}}\right)
\end{equation}

where:

$\tau$ - double pulse resolution  

$T_{acq}$ - acquisition time of single frame (GTU - dead time)

EUSO-TA detector can work with external trigger or can be self-triggered with 1Hz trigger rate.
While receiving trigger, 128 GTU frames (one packet) is saved, i.e. 64 frames before trigger, and 64 after.
An average trigger rate for work with TA is around 2Hz so stars are in almost the same position on focal surface between neighbouring packets.



\section{Atmospheric attenuation and predicted star signal}


To perform absolute calibration we need to know the parameters of the signal reaching EUSO detector.
In case of star, intensity of the signal depends on its spectrum and path in the atmosphere due to the Rayleigh and Mie scattering as well as the ozone absorption.
For calculations we used the Pickles stellar spectral flux library \cite{Pickles:1998dm}.
Spectra in the database are related to specific spectral types of stars therefore must be re-normalized for each case during analysis.
Atmospheric transmittance has been calculated using libRadTran package \cite{gmd-9-1647-2016} applying atmospheric parameters corresponding to the desert conditions where EUSO-TA detector is placed.
An accurate representation of atmospheric conditions during measurements is difficult task due to variability of aerosols.
For calculations we assumed average values of visibility.

\begin{figure}[!h]
\begin{center}
\includegraphics[width=1.00\textwidth]{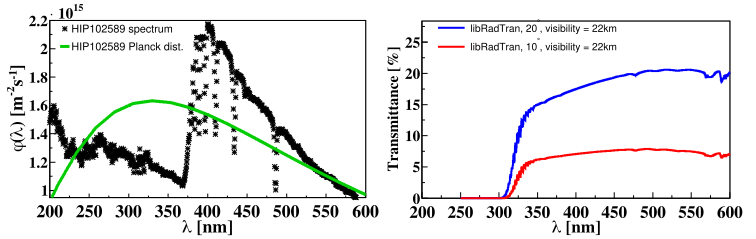}
	\caption{Comparison of photons spectrum for HIP102589 star with related Planck distribution at the top of the atmosphere (left) and atmospheric transmittance for observations at 10$^\circ$ and 20$^\circ$ of elevation, calculated with libRadTran.}
\label{libRadTran}
\end{center}
\end{figure}


Expected signal $N_{exp}$ for fixed observation angle can be simply described by following formula:
\begin{equation}
	N_{exp} = C_{abs}\int_{0}^{\infty}T_{atm}(\lambda)\Phi_{top}(\lambda)P_{det}(\lambda)d\lambda
\label{SignalFormula}
\end{equation}
\newpage
where:

$T_{atm}(\lambda)$ - atmospheric transmittance

$\Phi_{top}(\lambda)$ - star light flux at the top of the atmosphere

$P_{det}(\lambda)$ - known detector parameters

$C_{abs}$ - wavelength independent absolute calibration constant



\section{The data analysis}
\begin{figure}[!h]
\begin{center}
\includegraphics[width=.85\textwidth]{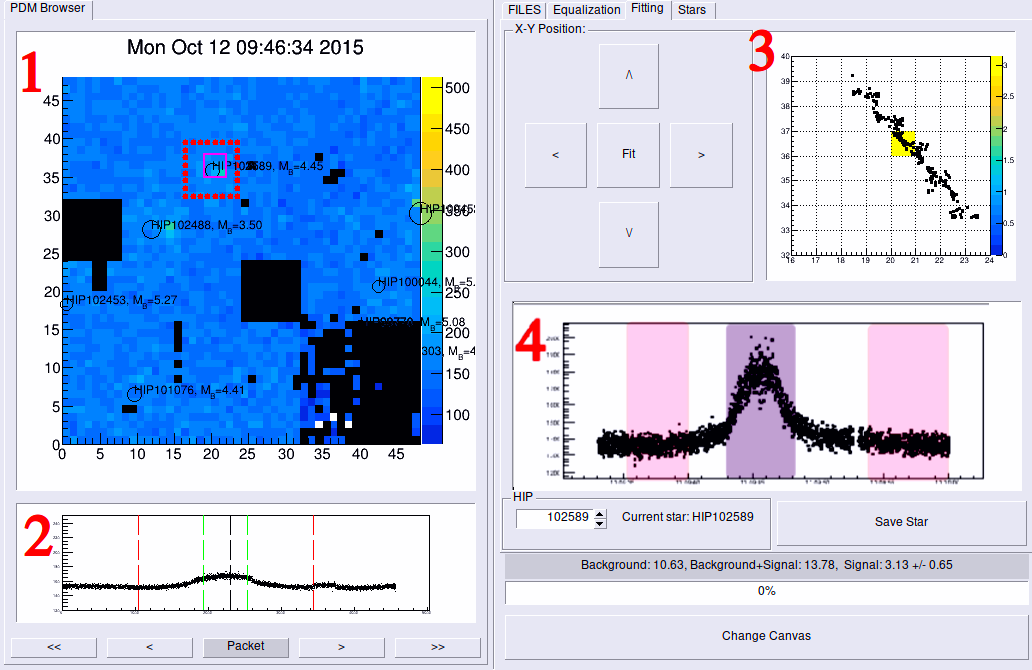}
\caption{The main panel of ROOT-based software dedicated for analysis of signals from stars in EUSO-TA data.}
\label{RootBrowser}
\end{center}
\end{figure}

Data analysis has been performed using dedicated GUI browser prepared in ROOT framework \cite{Brun:1997pa}.
The main panel is presented in Fig. \ref{RootBrowser}. 
We work on counts summarized over the packets.
Because of different response of individual pixels in the first step we use the equalization algorithm \cite{Plebaniak:2020hkn}.
This allows to fit the tracks of stars with M$_{B}$<$5.5^{m}$.
Pixels excluded by the algorithm due to malfunctioning or nonlinear response are marked as a black squares.
To determine stars in field of view we use the HIPPARCOS catalogue \cite{Perryman:1997sa}. 
Positions of few stars are displayed in canvas (1). 
Then, the light-curve representing the counts in entire PMT (2) is analyzed to find the range for fitting the track (3). 
In the next step we analyze the lighcurve for 3x3 pixels area around the pixel which star is going through (4). 
In this way we can determine background and signal region and extract the information about number of counts and its discrepancy.
All the process is manual and user have to take the decision if accuracy of the signal is good enough to classify it as an useful result or not.
This allow e.g. to exclude signals registered together with planes or lightnings in field of view.

\subsection{Results}
\begin{figure}[!h]
\begin{center}
\hspace*{-1.2cm}
\includegraphics[width=1.1\textwidth,height=0.9\textwidth]{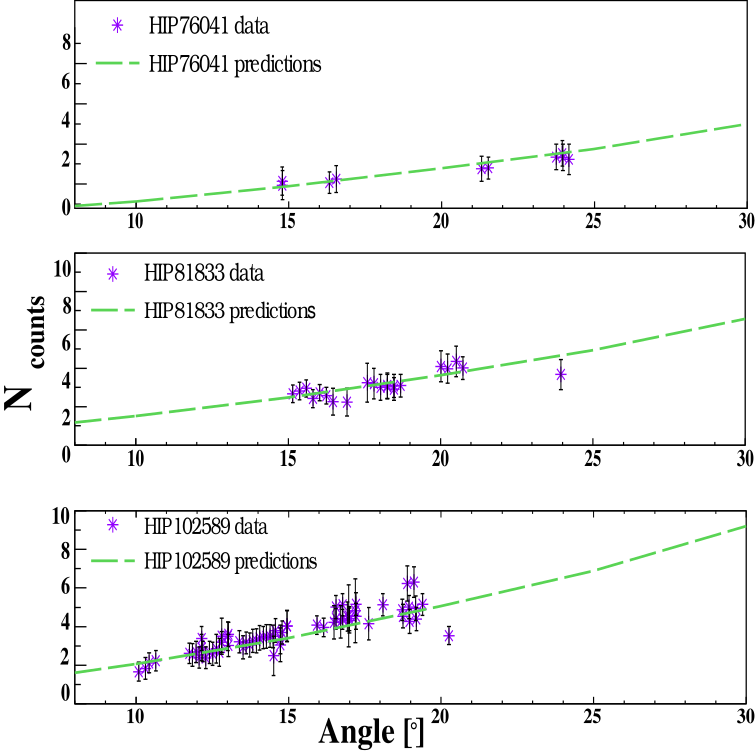}
\caption{
Comparison of measured and expected signals for three selected stars obtained in wide range of elevations angles.
All lines representing expected signal have been calculated using the same absolute calibration constant $C_{abs}$}
\label{DataVsPredictions}
\end{center}
\end{figure}

Following the procedure described above we analyzed signals from about 100 stars in the data taken by EUSO-TA in the years 2015 and 2016.
Some stars had to be removed from analysis because of their variability or being in the binary systems.
At the end, we get the list of several stars which can be treated as a point-like calibration sources for EUSO detectors.
Taking into account all of them, we determined the calibration constant in Eq. \ref{SignalFormula} as $C_{abs}$ = 0.86.
Table \ref{RegStars} contains the comparison of calculated expected and measured signal for several stars with different spectral types.
In the Fig. \ref{DataVsPredictions} we present an example of the comparison of predicted and measured number of photo-electrons for three stars in wide range of elevation angle.
An agreement at various angles confirms that atmospheric extinction model has been used correctly.
Assuming that obtained calibration constant should be taken into account as a detector parameter, we calculated the total detector efficiency for registration of point-like signal in 3x3 pixels area as 3.48, 5.81 and 5.60$\%$ for	300, 365 and 400 nm respectively.

\begin{center}

\begin{table}[!h]
\centering
\caption{Results comparison for several measured stars with various spectral types. \label{RegStars}}
\centering
\resizebox{1.00\textwidth}{!}{\begin{minipage}{\textwidth}
\begin{tabular}{|c|c|c|c|c|c|}
\hline
HIP & M$_B$ & Spectral Type & Angle & Measured N$_{ph}$/GTU & Expected N$_{ph}$/GTU \\
\hline
\hline
102488 & 2.44 & K0III & 12.31 & 5.93$\pm$0.99 &  6.06 \\
102589 & 4.54 & B5V & 14.73 & 2.96$\pm$0.85 & 3.21 \\
100453 & 2.23 & F8I& 14.55 & 11.40$\pm$1.58 & 11.89 \\
 50801 & 3.05 & M0III & 23.00 & 2.91$\pm$0.57 & 3.13 \\
 93194 & 3.25 & B9III & 11.71 & 8.20$\pm$1.40 & 7.97 \\
 81833 & 3.50 & G7III & 15.37 & 2.82$\pm$0.46 & 2.52 \\
 93903 & 5.25 & B6IV & 15.93 & 1.57$\pm$0.37 & 1.67 \\
109410 & 4.29 & F5III & 13.51 & 2.26$\pm$0.63 & 2.42 \\
 76041 & 4.98 & A2V & 14.79 & 2.15$\pm$0.70 & 1.82 \\
  4436 & 3.87 & A6V & 16.33 & 5.02$\pm$0.84 & 4.86 \\
\hline
\end{tabular}
\end{minipage}}
\end{table}

\end{center}

\vspace{-0.5cm}
\section{Conclusions}
Based on EUSO-TA data we have developed the procedure dedicated for data analysis of point-like sources, resulting absolute calibration of EUSO detector.
In the paper results of analysis of signals generated by stars have been presented.
For this case, we applied the model of atmospheric transmission to calculate expected signal and determine calibration constant of EUSO-TA detector.
The same procedure can be used for calibration of others EUSO detectors during on-ground tests as well as during measurements from stratospheric balloons or in space.
Scanning of the focal surface by signals induced by stars gives also an opportunity to monitor the conditions of individual pixels and optical system during observations of ultra high energy cosmic rays. 
\vspace{0.5cm}
\newline 
\newpage
{\small{\bf Acknowledgments:}
This work was partially supported by Basic Science Interdisciplinary Research
Projects of RIKEN and JSPS KAKENHI Grant (JP17H02905, JP16H02426 and
JP16H16737), by the Italian Ministry of Foreign Affairs and International
Cooperation, by the Italian Space Agency through the ASI INFN agreement
n. 2017-8-H.0 and contracts n. 2016-1-U.0 and n. 2021-8-HH.0, by ASI INAF
agreeement
n.2017-14-H.O, by NASA award 11-APRA-0058 in the USA, by the
Deutsches Zentrum f\"ur Luft- und Raumfahrt, by the French space agency
CNES, the Helmholtz Alliance for Astroparticle Physics funded by the
Initiative and Networking Fund of the Helmholtz Association (Germany), by
Slovak Academy of Sciences MVTS JEMEUSO as well as VEGA grant agency project
2/0132/17,
by National Science Centre in Poland grant no 2017/27/B/ST9/02162, by Mexican funding agencies PAPIIT-UNAM, CONACyT and
the
Mexican Space Agency (AEM).
Russian team is supported by ROSCOSMOS, "KLYPVE" is included into the
Long-term program of Experiments on board the Russian Segment of the ISS.
Sweden is funded by the Olle Engkvist
Byggm\"astare Foundation.

}


\bibliography{Bibliography}


\newpage
\noindent
{\Large\bf Full Authors list: The JEM-EUSO Collaboration\\}

\begin{sloppypar}
{\small \noindent 
G.~Abdellaoui$^{ah}$, 
S.~Abe$^{fq}$, 
J.H.~Adams Jr.$^{pd}$, 
D.~Allard$^{cb}$, 
G.~Alonso$^{md}$, 
L.~Anchordoqui$^{pe}$,
A.~Anzalone$^{eh,ed}$, 
E.~Arnone$^{ek,el}$,
K.~Asano$^{fe}$,
R.~Attallah$^{ac}$
% ...
}
\end{sloppypar}
\vspace*{.3cm}

{ \footnotesize
\noindent
% ...
$^{ac}$ LPR at Dept. of Physics, Faculty of Sciences, University Badji Mokhtar, Annaba, Algeria \\
% ...
$^{ah}$ Telecom Lab., Faculty of Technology, University Abou Bekr Belkaid, Tlemcen, Algeria \\
% ...
$^{cb}$ Universit\'e de Paris, CNRS, AstroParticule et Cosmologie, F-75013 Paris, France\\
% ...
$^{ed}$ Istituto Nazionale di Fisica Nucleare - Sezione di Catania, Italy\\
% ...
$^{eh}$ INAF - Istituto di Astrofisica Spaziale e Fisica Cosmica di Palermo, Italy\\
% ...
$^{ek}$ Istituto Nazionale di Fisica Nucleare - Sezione di Torino, Italy\\
$^{el}$ Dipartimento di Fisica, Universita' di Torino, Italy\\
% ...
$^{fe}$ Institute for Cosmic Ray Research, University of Tokyo, Kashiwa, Japan\\ 
% ...
$^{fq}$ Nihon University Chiyoda, Tokyo, Japan\\ 
% ...
$^{md}$ Universidad Polit\'ecnia de madrid (UPM), Madrid, Spain\\
% ...
$^{pd}$ University of Alabama in Huntsville, Huntsville, AL; USA\\
$^{pe}$ Lehman College, City University of New York (CUNY), NY, USA\\
% ...
}



\vspace*{0.5cm}

\vspace*{0.5cm}

\end{document}